\makeatletter
\let\old@footnotetext=\@footnotetext
\makeatother
\documentclass{aastex}
\makeatletter
\let\@footnotetext=\old@footnotetext
\makeatother
\usepackage{emulateapj5}

\newcommand{\chandra}{{\it Chandra}}
\newcommand{\xmm}{{\it XMM-Newton}}

\newcommand{\co}{CXOU~J112439.1$-$591620}
\newcommand{\gtwo}{G292.0+1.8}
\newcommand{\msh}{MSH~11$-$5{\sl 4}}
\newcommand{\psr}{PSR~J1124$-$5916}
\newcommand{\lsim}{\hbox{\raise.35ex\rlap{$<$}\lower.6ex\hbox{$\sim$}\ }}
\newcommand{\gsim}{\hbox{\raise.35ex\rlap{$>$}\lower.6ex\hbox{$\sim$}\ }}

\shorttitle{X-Ray Pulsar in SNR \gtwo}
\shortauthors{Hughes \& Slane}

\submitted{Received \underbar{~~~~~~~~~~~~}; accepted \underbar{~~~~~~~~~~~~}}
\accepted{Draft version of 20 May 2003}

\begin{document}

\title{An X-Ray Pulsar in the Oxygen-Rich Supernova Remnant \gtwo}

\author{
 John P.~Hughes\altaffilmark{1}, 
 Patrick O.~Slane\altaffilmark{2},
 Sangwook Park\altaffilmark{3},
 Peter W. A. Roming\altaffilmark{3},
 and
 David N. Burrows\altaffilmark{3}
}
\altaffiltext{1}{Department of Physics and Astronomy, Rutgers
University, 136 Frelinghuysen Road, Piscataway, NJ 08854-8019;
jph@physics.rutgers.edu}
\altaffiltext{2}{Harvard-Smithsonian Center for Astrophysics, 60 Garden Street,
Cambridge, MA 02138; slane@head-cfa.harvard.edu}
\altaffiltext{3}{Department of Astronomy and Astrophysics, Pennsylvania State
University, 525 Davey Laboratory, University Park, PA. 16802}

\begin{abstract}

We report the discovery of pulsed X-ray emission from the compact
object \co\ within the supernova remnant (SNR) \gtwo\ using the High
Resolution Camera on the {\it Chandra X-ray Observatory}.  The X-ray
period ($P = 0.13530915$ s) is consistent with extrapolation of the
radio pulse period of \psr\ for a spindown rate of $\dot P = 7.6\times
10^{-13}$ s/s. The X-ray pulse is single peaked and broad with a FWHM
width of 0.23$P$ (83$^\circ$).  The pulse-averaged X-ray spectral
properties of the pulsar are well described by a featureless power law
model with an absorbing column density $N_{\rm H} = 3.1\times 10^{21}$
cm$^{-2}$, photon index $\Gamma = 1.6$, and unabsorbed 0.3--10 keV
band luminosity $L_X = 7.2\times 10^{32}$ erg s$^{-1}$.  We plausibly
identify the location of the pulsar's termination shock.  Pressure
balance between the pulsar wind and the larger synchrotron nebula, as
well as lifetime issues for the X-ray-emitting electrons, argues for a
particle-dominated PWN that is far from the minimum energy condition.
Upper limits on the surface temperature of the neutron star are at, or
slightly below, values expected from ``standard'' cooling
curves. There is no optical counterpart to the new pulsar; its optical
luminosity is at least a factor of 5 below that of the Crab pulsar.

\end{abstract}

\keywords{ ISM: individual (SNR \gtwo, \msh) -- pulsars: individual
(\psr) -- stars: neutron -- supernova remnants -- X-rays: individual
(\co) }

\section{Introduction}

The composition of the ejecta seen in a supernova remnant (SNR) can be
used to constrain the nature of the supernova (e.g., core collapse or
thermonuclear) and, in the case of a core collapse SN, estimate a
range for the mass of the progenitor star (e.g., Hughes \& Singh
1994). Recent studies of the SNR Cas A with \chandra\ and \xmm\ (e.g., Hughes
et al.~2000; Bleeker et al.~2001; Willingale et al.~2002) highlight
the great potential of the new observatories for such studies.
Unfortunately, most SNRs that harbor young pulsars are virtually
useless for such investigations: they do not show much evidence for
shocked ejecta (e.g., the Crab and 3C 58), are too distant for
detailed study (e.g., SNR E0540$-$69.3 and N157B), or are so evolved
that the ejecta cannot be easily distinguished from shocked
interstellar gas (e.g., Vela and W44).  \gtwo\ differs from nearly all
other pulsar/SNR associations by virtue of showing spectacular
evidence for newly synthesized oxygen-, neon-, and magnesium-rich
ejecta (optical: Murdin \& Clark 1979; X-ray: Park et al.~2002);
having a dynamically determined age ($\sim$2000 yrs; Murdin \& Clark
1979); and being relatively nearby ($\sim$6 kpc; Gaensler \& Wallace
2003).  With \gtwo\ we have the opportunity to tie information on the
progenitor star derived from nucleosynthesis (Park et al.~2003, in
prep.)  to the origin and evolution of pulsars and their wind nebulae
(PWNe). 

Recently \chandra\ revealed an X-ray point source (\co) centered on a
diffuse synchrotron nebula in \gtwo\ (Hughes et al.~2001). In a
follow-up study Camilo et al.~(2002) discovered a 135-ms young pulsar
within or near \gtwo\ using the Parkes radio telescope that is almost
surely the counterpart to the \chandra\ point source.  However the
large beam of the Parkes telescope ($\sim$14$^\prime$ FWHM) means that
the case is not ironclad.  Here we report on high temporal and spatial
resolution X-ray observations in which we detect the pulsed signal
from \co, clearly identifying it as the compact remnant of the SN that
formed \gtwo.

\section{X-ray Pulsar}

We observed SNR \gtwo\ beginning on 14 July 2001 using the \chandra\
High Resolution Camera (HRC) in timing mode with the pulsar candidate
at the aimpoint (ObsID 1953). Timing mode observations utilize the
central portion of the HRC-S focal plane array which provides a field
of view roughly 6$^\prime$ by 30$^\prime$. Individual photons are time
tagged to an accuracy of about 16 $\mu$s; we corrected photon arrival
times to the solar-system barycenter using the position of the pulsar
candidate. Our observation was nearly continuous; the only
interruptions were 8 gaps of $\sim$2 s duration each, distributed
throughout the exposure. The livetime corrected exposure was 49578 s.

\begin{figure*}[t]
\vspace{-0.5truein}
\begin{center}
\epsfxsize=6.5truein\epsfbox{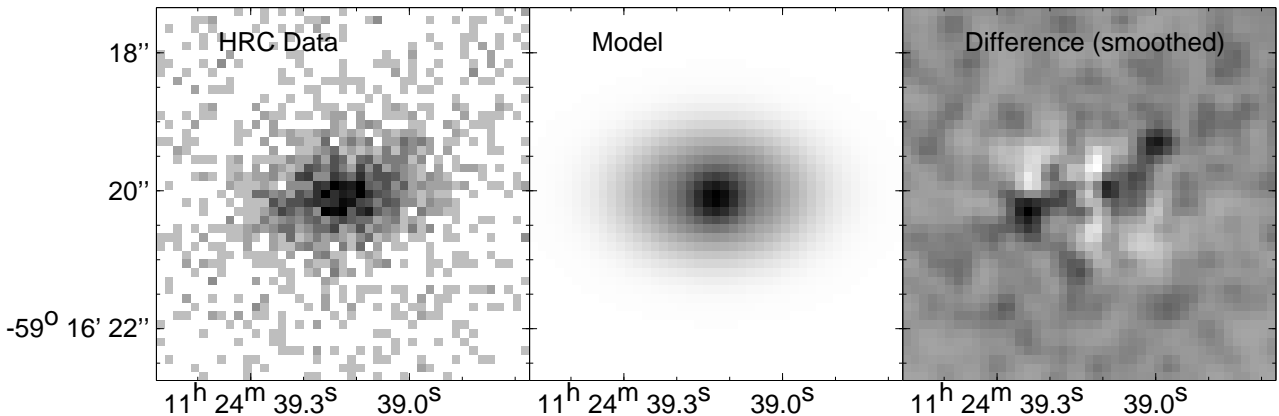}
\end{center}
\vspace{-3.85truein}
\figcaption{
A portion of the \chandra\ high resolution
camera image centered on \co. In the left and middle
panels the grayscale is linear from 0 to 15 HRC counts per pixel
(0.1318$^{\prime\prime}$ square).  In the right panel the linear
grayscale extends from -1.4 to 1.8 counts per pixel. This last image
was smoothed with a 1 pixel $\sigma$ gaussian kernel. Coordinates
are given in epoch J2000.
}
\end{figure*}

Figure 1 (left panel) shows a roughly $6^{\prime\prime} \times
6^{\prime\prime}$ portion of the HRC image containing the pulsar
candidate at position R.A.~=~11:24:39.1, decl.~=~$-$59:16:20 (J2000).
There is an unresolved source centered on a small, diffuse,
elliptically-shaped nebula. Within a radius of 2$^{\prime\prime}$ (15
HRC pixels) of the point source we detected 1324 X-ray photons. First
we carried out a blind search for pulsations on these events.  Light
curves were constructed for the entire duration of the observation
using four different binsizes of 0.0237563 s, 0.0118781 s, 0.00593907
s, and 0.00296954 s corresponding to $2^{21}$, $2^{22}$, $2^{23}$, and
$2^{24}$ temporal bins. A coherent FFT of the entire light curve
showed no statistically significant pulsed signal for any of these
cases.  The distribution of Fourier powers was consistent with noise
and the individual peak Fourier powers obtained were 30.4, 31.9, 32.9,
and 34.2 for the four cases, respectively, none of which are
statistically significant. As a verification of our methods and IDL
software we applied the same programs to the HRC data of PSR
B0540$-$69.3, observed on 22 June 2000 (ObsID 1745) using the same
configuration as our data.  The pulsar was easily detected at a
frequency of 19.7941 Hz with a peak Fourier power of 50.3 (99.998\%
significance).

A much more sensitive search for X-ray pulsations is possible by
narrowing the range of trial frequencies to be consistent with the
radio pulse and a reasonable range of $\dot P$ values. We employed the
$Z_n^2$ test (Buccheri et al.~1983) which applies a harmonic analysis
to the phases of photon arrival times for a given trial pulsation
frequency.  One advantage of the method, compared to epoch-folding for
example, is that it requires no binning. Another is that, even for as
few as 100 detected photons, the statistic is distributed like $\chi^2$
with $2n$ degrees of freedom.  In our searches we use $n=2$.

We searched eleven trial frequencies spaced by $\Delta f = 1\times
10^{-5}$ Hz (roughly the frequency resolution of our data) and
centered on the expected value based on extrapolating the radio
ephemeris to the midpoint of the HRC observation (MJD = 52105.18). The
peak $Z_2^2$ value was 22.6 corresponding to the 99.8\% significance
level. The search was refined by reducing $\Delta f$ to $2\times
10^{-6}$ Hz and again searching eleven trial frequencies, this time
centered on the most likely previous pulsation frequency. This
iteration yielded a peak $Z_2^2$ value of 27.9 (99.97\% significant or
approximately 3.6 $\sigma$) at a period of 0.13530915 s.  The period
error ($4\times 10^{-8}$ s, 1 $\sigma$) was determined using a
bootstrap algorithm. In Table 1 we quote observed properties of the
X-ray pulsar.  By comparing our pulse period to the value obtained by
Camilo et al.~(2002) roughly two months later we derive a period
derivative of $\dot P = 7.62\pm0.06\times 10^{-13}$ s/s that differs
by $\sim$2.5 $\sigma$ from the value quoted in the radio discovery
paper.  At present we do not know the relative X-ray and radio pulse
phases and, due to apparent rotational instabilities in the neutron
star, it is not possible to extrapolate the radio ephemeris from
September 2001 back to July 2001 accurately enough to measure relative
phases.

As an additional check on the detection of pulsed X-ray emission, we
applied the last search iteration to the first and second halves of
the data set (split in time) independently.  The pulse was detected in
each half at the appropriate $Z_2^2$ value and pulsation frequency and
with similar light curve shapes.

\begin{center}
\begin{minipage}[t]{0.47\textwidth} 
\epsfxsize=0.98\textwidth \epsfbox{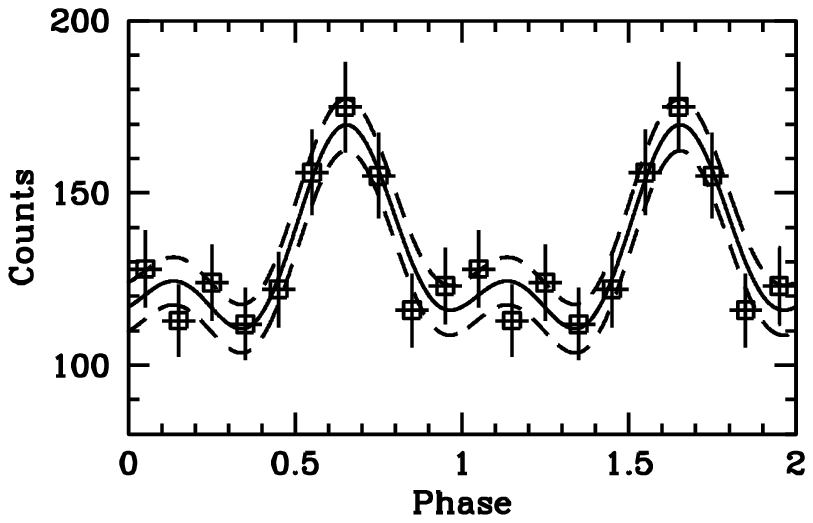}
\figcaption{
Pulse phase light curve for \psr\ folded modulu the
best-fit period of 0.13530915 s. Two complete periods are shown. Note
the suppressed zero on the y-axis. Also plotted are the Fourier series
estimator (de Jager, Swanepoel, \& Raubenheimer 1986) of the light
curve (solid curve) and its 1 $\sigma$ uncertainty (dashed curves).  }
\end{minipage}
\end{center}

The pulse in the X-ray band is single peaked and symmetric (see
Fig.~2), similar to the radio pulse, although the X-ray pulse width
(FWHM $\sim 0.23P \sim 83^\circ$) is somewhat broader than the radio
one.  The smooth curve in figure 2 is a Fourier series estimate (de
Jager, Swanepoel, \& Raubenheimer 1986) of the light curve employing
two harmonics. If we assume the pulse extends over phase bins
0.43--0.90, we determine the fraction of pulsed X-rays in the
2$^{\prime\prime}$ radius extraction region to be $11\pm1$\%.  This
includes contribution from the diffuse compact nebula, which we
quantify next.

{\parindent=0cm 
\begin{small}
\begin{minipage}[t]{0.47\textwidth}
\begin{center}
{\noindent{TABLE 1}}\\
{\noindent{\sc Properties of X-ray \psr}}\\[6pt]
\begin{tabular}{@{}lcc@{}}
\hline\hline\\[-4pt]
Parameter & Value\\[4pt]
\hline\\[-4pt]
R.A. (J2000)          &  11 24 39.1     \\[2pt]
Decl. (J2000)         &  $-$59 16 20    \\[2pt]
Period, $P$ (s)       &  0.13530915(4)  \\[2pt]
Epoch (MJD)           &  52105.18       \\[2pt]
Observation span (hr) &  14.3           \\[2pt]
FWHM of pulse         &  $\sim$0.23$P$  \\[2pt]
Pulsed fraction (\%)  &  $91^{+\phantom{2}9}_{-24}$ \\[2pt]
HRC rate (s$^{-1}$)   &  0.0032(8)      \\[2pt]
Column density, $N_{\rm H}$ (cm$^{-2}$)  & $3.1(4)\times 10^{21}$ \\[2pt]
Photon index, $\Gamma$ & 1.6(1)            \\[2pt]
Luminosity, $L_X$(0.3--10 keV) (erg s$^{-1}$) &  $7.2 \times 10^{32}\, 
(D/6\,\rm  kpc)^2$ \\[2pt]
\quad (Unabsorbed) & \\[2pt]
\hline\\[-8pt]
\end{tabular}
\end{center}
Note.---Numbers in parentheses represent 1 $\sigma$ uncertainties in
the least significant digits quoted.
\end{minipage}
\end{small}
}

\section{Extended Compact Nebula}

Shown in the middle panel of figure 1 is our best fit spatial model
for the HRC data: an unresolved point source (i.e., a gaussian whose
best-fit angular size is consistent with the \chandra\ PSF) and an
elliptical gaussian with a FWHM of 1.8$^{\prime\prime}$ (along the
major axis) and an axial ratio of 2.  In this model the point source
contains $160\pm40$ X-ray events while the extended elliptical
component contains 1440 events.  Compared to the number of pulsed
events we detect ($146\pm13$), it is clear that the point source
itself is highly pulsed with a pulsed fraction of $>$65\% in the
HRC band.  

The rightmost panel in Figure 1 shows the difference between the HRC
data and the best-fit image model. There is good evidence for excess
X-ray emission above that given by the model, to either side of the
point source and oriented generally in the SE-NW direction.  One
possibility is that the excess emission comes from a pair of jets.
This feature is nearly aligned with the direction from the current
position of the point source back toward the center of the SNR (toward
the NW), which would indicate aligned spin axis and proper motion
directions for \psr, as seen in the Crab and Vela pulsars. On the
other hand, it is also possible that the excess emission arises from a
toroidal structure in the nebula (like the torus in the Crab Nebula)
seen in projection. In this scenario the torus would be nearly aligned
with the major axis of the compact nebula.  By analogy to the Crab and
its pulsar, we would therefore expect that the spin axis to be
perpendicular to the long axis of the compact nebula (i.e., aligned
NE-SW).  This would put the pulsar's spin axis nearly perpendicular to
its proper motion direction.  The current \chandra\ data do not allow
us to discriminate between these possibilities.

The extended compact nebula is the only emission feature in the PWN
within an arcmin or so of the pulsar and therefore is the only plausible
candidate for the pulsar wind termination shock.  If we interpret the
edge of the nebula with the location of this shock, we can then
estimate the confining pressure (i.e., in the PWN) necessary to
balance the ram pressure of the wind, $P_{\rm w} = \dot E / 4\pi c
r_{\rm w}^2$, assuming spherical symmetry.  The mean radius of the
compact nebula is $0.036\,d_6$ pc and the spin down energy loss of the
pulsar is $\dot E = 1.2\times 10^{37}$ erg s$^{-1}$ (Camilo et al.~2002),
so the pressure is $P_{\rm w} = 2.6\times 10^{-9} \,d_6^{-2}$ erg
cm$^{-3}$.

We estimate the pressure in the PWN from the properties of the radio
emission, which we take from Gaensler \& Wallace (2003), and the
theory of synchrotron emission (Longair 1994).  Under the minimum
energy condition we find that $P_{\rm PWN,min} \sim 1.3\times
10^{-10}\,d_6^{-4/7}$ erg cm$^{-3}$ assuming equal energy densities in
the protons and electrons and a volume filling factor of unity.  The
average nebular magnetic field under these conditions is $B_{\rm min}
\sim 48 \,d_6^{-2/7}$ $\mu$G, which implies a very short synchrotron
lifetime, $t \sim 140\, d_6^{3/7}\, (h\nu/ 2 \,\rm keV)^{-1/2}$ yr,
for the electrons giving rise to the X-ray emission. Such a short
lifetime is inconsistent with the observation that the X-ray
synchrotron nebula covers as large an extent as the radio nebula does.

A possible solution to these discrepancies lies in relaxing the
minimum energy condition.  If we move in the direction of a smaller
mean nebular magnetic field we resolve the lifetime issue.  A magnetic
field strength of $\lsim$8$\mu$G would ensure that the X-ray
synchrotron cooling time is $\gsim$2000 yr.  In order that the
pressure in the synchrotron nebula be sufficiently strong to balance
the ram pressure of the pulsar's wind requires a value of $B\sim3$
$\mu$G.  The total energy in the nebula would then be $\sim$$4\times
10^{49}$ ergs, contained nearly entirely in particles. Since this
energy has come from the spin-down of the pulsar, it sets a constraint
on the initial spin period: $P_0 \sim 22$ ms for a canonical NS
momentum of inertia of $I \equiv 10^{45}$ g cm$^2$. This $P_0$ value
is considerably less than the value of $\sim$90 ms estimated by Camilo
et al.~(2002).  The simplest way to accommodate our low value for the
initial spin period would be to increase the true age of the pulsar to
$\sim$2800 yr or more.

It is important to note that neither the magnetic field nor the
pressure is expected to be uniform in PWNe, as we assumed in the
calculations above.  In the Kennel \& Coroniti (1984) model for the
Crab Nebula the total pressure is greatest at the termination shock
and then falls by factors of 3--10 at larger radii. In addition,
equipartition between particles and fields is attained only at a
significant distance from the pulsar; near the termination shock the
magnetic field is low and the pressure is particle-dominated.  Because
of the higher central pressure, we expect the volume-averaged magnetic
field under this model to be somewhat larger than that estimated
above, which would have the effect of relaxing the energetics
constraint on the pulsar's initial spin period.  Our apparent need for
a particle-dominated PWN in \gtwo\ is suggestive of a low value for
the magnetization parameter in the context of this model. We note,
however, that interaction between the reverse shock and the PWN (which
has not yet been conclusively established) may offer an alternate
explanation for why the nebula is far from the minimum energy
condition.  Further study of these issues, although beyond the scope
of our work here, is clearly warranted.

\section{Neutron Star Cooling}

The NS in \gtwo\ is quite young with a most likely age range of
$\sim$2000 yrs to $\sim$2900 yrs, corresponding to the free expansion
age of the O-rich knots and the pulsar characteristic age, respectively.
According to NS cooling models (e.g., Tsuruta 1998; Page 1998), the
surface temperature at this age should be high enough to produce
detectable X-ray emission. As
shown by Hughes et al.~(2001), the ACIS-S spectrum of the pulsar is fully
consistent with a single absorbed power-law. Here we determine the
upper limit to the intensity of an additional blackbody spectral
component as a function of its temperature, $T_{BB}$.  We utilized two
independent spectra extracted from the CTI-corrected data (Park et
al.~2002): one from a 3$\times$3 pixel
(1.5$^{\prime\prime}$$\times$1.5$^{\prime\prime}$) region centered on
the pulsar, and another, comprising the diffuse nebula, from an ellipse
of size 7$\times$11 pixels (3.4$^{\prime\prime}$$\times$
5.4$^{\prime\prime}$) excluding the central pulsar region.  The pulsar
spectrum was fit to the sum of a blackbody and a power-law model
including absorption, while the nebular spectrum was fit to an
absorbed power-law model alone.  This latter spectrum served as an
independent constraint on the column density, which was constrained to
be the same between the two spectra. For reference, table 1 lists pure
power-law spectral parameters for the pulsar.

For a given fixed value of $T_{BB}$, the ACIS-S data set an upper
limit on the allowed normalization (or flux) of the blackbody
component.  One can express the normalization limit in terms of the
square of the ratio of the blackbody emitter's radius to its distance.
The 3 $\sigma$ limit on this ratio as a function of $T_{BB}$ is
plotted in figure 3.  

\begin{center}
\begin{minipage}[t]{0.47\textwidth} 
\epsfxsize=0.98\textwidth \epsfbox{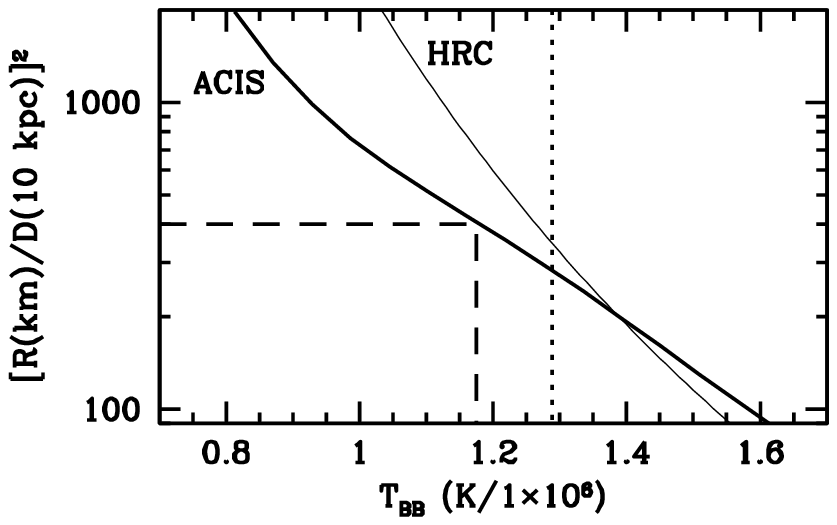}
\figcaption{
Constraint on the normalization of a blackbody spectral 
component vs.\ its temperature from fits to the time-averaged ACIS-S
spectrum of PSR J1124$-$5916 (thick solid curve).  The thin solid
curve indicates the constraint based on the unpulsed HRC count rate of
the X-ray pulsar.  The allowed region lies to the left and below the
curves shown.  The dashed lines show the temperature constraint for
the nominal value of distance to \gtwo (6 kpc) and a 12 km radius NS.
The vertical dotted line shows the temperature expected for a standard
NS cooling curve.}
\end{minipage}
\end{center}

Since the ACIS-S spectrum is consistent with an entirely nonthermal
origin, the pulsed emission seen in the HRC, which comprises
$>$65\% of the total HRC rate from the pulsar, therefore must be
dominated by nonthermal, i.e., magnetospheric, emission as well.  The
unpulsed HRC emission, however, can be used to set another constraint
on the mean surface temperature of the NS.  We convert the 3 $\sigma$
upper limit on the unpulsed HRC count rate ($2.8\times 10^{-3}$
s$^{-1}$), assuming the 3 $\sigma$ upper limit on the column density
to the pulsar ($N_{\rm H} = 4.75\times 10^{21}$ atom cm$^{-2}$, derived
from the nebular spectrum) to a constraint on the blackbody
normalization as a function of $T_{BB}$.  This constraint, which is
fully consistent with the one from the ACIS-S spectral analysis,
is shown as the thin curve in figure 3.

Recent work (Gaensler \& Wallace 2003) suggests that the distance to
\gtwo\ is $\sim$6 kpc.  Using this value and assuming a 12 km radius for the
NS, we obtain a constraint of $T_{BB} < 1.18\times 10^6\,\rm K$ on the
surface temperature of the NS.  The expected temperature, assuming
standard NS cooling models, is $1.28\times 10^6\,\rm K$ (Page
1998). Although this is suggestive of the presence of exotic cooling
processes, systematic uncertainties make this result less secure than
the recent result on the apparent need for exotic cooling processes
for the NS in 3C 58 (Slane, Helfand, \& Murray 2002).  The NS in
\gtwo\ would be consistent with standard cooling if it were as distant
as 7 kpc, or if the compact star's radius were as small as 10 km.
On the other hand pure blackbody spectral models tend to overpredict
(by factors of 1.5 or more) the effective temperature of NS surfaces
when light element atmospheres are included (Lloyd, Hernquist, \& Heyl
2002).

\section{Limits on an Optical Counterpart}

Optical emission from isolated pulsars within supernova remnants has
currently been detected from only four objects: PSR B0531+21 (Crab),
PSR B0540$-$69.3 (in the LMC), PSR B1509$-$58 (G320.4$-$1.2), and PSR
B0833$-$45 (Vela) (see, for example, Nasuti et al.~1997 and references
therein).  The first three are very young pulsars (1000--2000 years
old), while the pulsar in Vela is considerably older ($\sim$10,000
yrs), although it is still rather young compared to the average radio
pulsar.  Across the optical band these pulsars show flat power-law
spectra ($\alpha \sim 0$ for $F_\nu \propto \nu^{-\alpha}$), although
their intrinsic luminosity densities (i.e., $L_\nu = 4\pi D^2F_\nu$)
span 5 orders of magnitude from 0.5--$2\times 10^{19} \, \rm erg\,
s^{-1}\, Hz^{-1}$ (PSR B0531+21, PSR B0540$-$69.3, and PSR B1509$-$58)
to 3--$6\times 10^{14} \, \rm erg\, s^{-1}\, Hz^{-1}$ (PSR
B0833$-$45). In terms of age and remnant optical properties (i.e., the
presence of high velocity, oxygen-rich optical emission), \gtwo\ most
closely resembles SNR 0540$-$69.3.  However in terms of spin-down
energy loss ($\sim$$10^{37}$ erg s$^{-1}$), the pulsar in \gtwo\ is
more similar to PSR B0833$-$45 and PSR B1509$-$58.

With \chandra\ we have localized the \gtwo\ pulsar to an absolute
position accuracy of $\sim$1$^{\prime\prime}$.  Within double this
error circle there is no optical counterpart visible in the Digitized
Sky Survey.  We have obtained an upper limit on optical emission from
the pulsar, $B \gtrsim 22$, based on a narrow-band blue continuum
image of \gtwo\ taken by P.F.\ Winkler and K.S.\ Long from the CTIO
4-m in 1991. This corresponds to an intrinsic luminosity density of
$L_\nu < 3\times 10^{18} \, \rm erg\, s^{-1}\, Hz^{-1}$ (assuming a
distance of 6 kpc and extinction of $A_{\rm B} \sim 2.3$).  This is
about an order of magnitude less than the optical emission of the Crab
and SNR 0540$-$69.3 pulsars, but is only about a factor of two less
than the optical emission from PSR B1509$-$58 (Caraveo, Mereghetti, \&
Bignami 1994). A considerably fainter upper limit, based on data
acquired at CTIO in April 2002, will be the subject of a forthcoming
article.

\acknowledgments

We are grateful to Fernando Camilo and Bryan Gaensler for sharing
results or data prior to publication and to Frank Winkler for
supplying the optical image of \gtwo.  Mike Juda gave us some helpful
advice regarding the HRC data. We thank Karen Lewis and John Nousek
for their help with the initial proposal for HRC time.  We also thank
Simon Johnston for his useful comments as referee.  Partial support
for this research was provided by \chandra\ grant GO1-2052X to JPH.

\end{document}